\documentclass[12pt,a4paper]{article}
\usepackage{amssymb}
\usepackage{amsmath}
\usepackage{amsfonts}
\usepackage{geometry}
\usepackage{setspace}
\usepackage{rotating}

\newtheorem{lemma}{Lemma}

\input{tcilatex}

\begin{document}

\title{Further results on the estimation of\linebreak dynamic panel logit
models with fixed effects.}
\author{Hugo Kruiniger\thanks{%
Address: hugo.kruiniger@durham.ac.uk; Department of Economics, 1 Mill Hill
Lane, Durham DH1 3LB, England. I thank N. Peyerimhoff for helpful comments.
All remaining errors are mine.} \\
Durham University}
\date{This version: 28 January 2023\\
Previous versions: 27 October 2020, 26 April 2021 and 17 January 2023 }
\maketitle

\vspace{6.2cm}

\bigskip

\bigskip

\bigskip

\noindent JEL\ classification: C12, C13, C23.\bigskip

\noindent Keywords: dynamic panel logit models, exogenous regressors, fixed
effects.

\setcounter{page}{0} \thispagestyle{empty}

\newpage

\baselineskip=20pt

\renewcommand{\baselinestretch}{1.5}

\begin{center}
\textbf{Abstract}
\end{center}

\vspace{1cm}

Kitazawa (2013, 2016) showed that the common parameters in the panel logit
AR(1) model with strictly exogenous covariates and fixed effects are
estimable at the root-n rate using the Generalized Method of Moments. Honor%
\'{e} and Weidner (2020) extended his results in various directions: they
found additional moment conditions for the logit AR(1) model and also
considered estimation of logit AR(p) models with $p>1$. In this note we
prove a conjecture in their paper and show that for given values of the
initial condition, the covariates and the common parameters $2^{T}-2T$ of
their moment functions for the logit AR(1) model are linearly independent
and span the set of valid moment functions, which is a $2^{T}-2T\,$%
-dimensional linear subspace of the $2^{T}$-dimensional vector space of real
valued functions over the outcomes $y\in \{0,1\}^{T}$. We also prove that
when $p=2$ and $T\in \{3,4,5\},$ there are, respectively, $2^{T}-4(T-1)$ and 
$2^{T}-(3T-2)$ linearly independent moment functions for the panel logit
AR(2) models with and without covariates.

\setcounter{page}{0} \thispagestyle{empty}\newpage

\section{Proof of a conjecture in Honor\'{e} and Weidner (2020)}

We adopt the notation of Honor\'{e} and Weidner (2020). In p.16 of their
paper they define for triples of time periods $t,s,r\in \{1,2,...,T\}$ with $%
t<s<r$ the moment functions $m_{y_{0}}^{(a/b)(t,s,r)}(y,x,\beta ,\gamma ).$
Let $z_{t,s}(y_{0},y,x,\beta ,\gamma )=(x_{t}-x_{s})^{\prime }\beta +\gamma
(y_{t-1}-y_{s-1}).$ Then%
\begin{eqnarray*}
m_{y_{0}}^{(a)(t,s,r)}(y,x,\beta ,\gamma ) &=&\left\{ 
\begin{array}{l}
\exp [z_{t,s}(y_{0},y,x,\beta ,\gamma )] \\ 
\exp [z_{t,r}(y_{0},y,x,\beta ,\gamma )] \\ 
-1 \\ 
\exp [z_{r,s}(y_{0},y,x,\beta ,\gamma )]-1 \\ 
0%
\end{array}%
\right. 
\begin{array}{l}
\text{if }(y_{t},y_{s},y_{r})=(0,1,0), \\ 
\text{if }(y_{t},y_{s},y_{r})=(0,1,1), \\ 
\text{if }(y_{t},y_{s})=(1,0), \\ 
\text{if }(y_{t},y_{s},y_{r})=(1,1,0), \\ 
\text{otherwise,}%
\end{array}
\\
m_{y_{0}}^{(b)(t,s,r)}(y,x,\beta ,\gamma ) &=&\left\{ 
\begin{array}{l}
\exp [z_{s,r}(y_{0},y,x,\beta ,\gamma )]-1 \\ 
-1 \\ 
\exp [z_{r,t}(y_{0},y,x,\beta ,\gamma )] \\ 
\exp [z_{s,t}(y_{0},y,x,\beta ,\gamma )] \\ 
0%
\end{array}%
\right. 
\begin{array}{l}
\text{if }(y_{t},y_{s},y_{r})=(0,0,1), \\ 
\text{if }(y_{t},y_{s})=(0,1), \\ 
\text{if }(y_{t},y_{s}.y_{r})=(1,0,0), \\ 
\text{if }(y_{t},y_{s},y_{r})=(1,0,1), \\ 
\text{otherwise.}%
\end{array}%
\end{eqnarray*}%
In p.17 of their paper they conjecture that for $\gamma _{0}\neq 0$ (and
arbitrary $y_{0}$, $x$ and $\beta _{0};$ index $i$ is omitted) any moment
function $m_{y_{0}}(y,x,\beta ,\gamma )=\overline{w}%
(y_{1},...,y_{t-1})m_{y_{0}}^{(a/b)(t,s,r)}(y,x,\beta ,\gamma )$ for the
panel logit AR(1) model with strictly exogenous regressors and $T\geq 3,$
where $\overline{w}_{y_{1},...,y_{t-1}}(y_{1},...,y_{t-1}):\{0,1\}^{t-1}%
\rightarrow 
\mathbb{R}
$, can be written as%
\begin{eqnarray*}
m_{y_{0}}(y,x,\beta ,\gamma )
&=&\sum\limits_{t=1}^{T-2}\sum%
\limits_{s=t+1}^{T-1}[w_{y_{0}}^{(a)}(t,s,y_{1},...,y_{t-1},x,\beta ,\gamma
)m_{y_{0}}^{(a)(t,s,T)}(y,x,\beta ,\gamma ) \\
&&+w_{y_{0}}^{(b)}(t,s,y_{1},...,y_{t-1},x,\beta ,\gamma
)m_{y_{0}}^{(b)(t,s,T)}(y,x,\beta ,\gamma )]
\end{eqnarray*}%
with weights $w_{y_{0}}^{(a/b)}(t,s,y_{1},...,y_{t-1},x,\beta ,\gamma )\in 
\mathbb{R}
$ that are uniquely determined by the function $m_{y_{0}}(.,x,\beta ,\gamma
).$

We will prove this conjecture by showing for given values of $y_{0},$ $x,$ $%
\beta $ and $\gamma $ (i) that the set of valid moment functions is a linear
subspace of the $2^{T}$-dimensional vector space of real valued functions
over the outcomes $y\in \{0,1\}^{T}$ that has a dimension of at most $%
2^{T}-2T,$ and (ii) that the $2^{T}-2T$ moment functions of the form $%
w_{y_{1},...,y_{t-1}}(y_{1},...,y_{t-1})m_{y_{0}}^{(a/b)(t,s,T)}(y,x,\beta
,\gamma ),$ where $w_{y_{1},...,y_{t-1}}(y_{1},...,y_{t-1}):\{0,1\}^{t-1}%
\rightarrow \{0,1\}$ are $2^{t-1}$ linearly independent indicator functions
and $1\leq t<s<T,$ are linearly independent and span this subspace.

Proof: Recall that $Pr(Y_{i}=y_{i}|Y_{i0}=y_{i0},X_{i}=x_{i},A_{i}=\alpha
_{i})\equiv $\pagebreak 
\begin{equation*}
p_{y_{i0}}(y_{i},x_{i},\beta _{0},\gamma _{0},\alpha
_{i})=\prod\limits_{t=1}^{T}\frac{1}{1+exp[(1-2y_{it})(x_{it}^{\prime }\beta
_{0}+y_{i,t-1}\gamma _{0}+\alpha _{i})]}.
\end{equation*}

We drop the index $i.$ A valid moment function $m_{y_{0}}(y,x,\beta ,\gamma
) $ satisfies%
\begin{equation*}
E[m_{y_{0}}(Y,X,\beta _{0},\gamma _{0})|Y_{0}=y_{0},X=x,A=\alpha ]=0\text{
for all }\alpha \in 
\mathbb{R}%
\end{equation*}%
or equivalently%
\begin{equation*}
\sum\limits_{y\in \{0,1\}^{T}}p_{y_{0}}(y,x,\beta _{0},\gamma _{0},\alpha
)m_{y_{0}}(y,x,\beta _{0},\gamma _{0})=0\text{ for all }\alpha \in 
\mathbb{R}
.
\end{equation*}

Let $T\geq 2$ and $\alpha _{1}<\alpha _{2}<...<\alpha _{2^{T}}.$ Define the $%
2^{T}\times 2^{T}$ matrix $\bar{P}$ with $\bar{P}_{g,h}=p_{y_{0}}(y,x,\beta
_{0},\gamma _{0},\alpha _{g})$ for $g,h=1,2,...,2^{T}$ with $%
h=1+2^{0}y_{1}+2^{1}y_{2}+...+2^{T-1}y_{T}.$ Next let $P_{g,t}=\exp
(x_{t}^{\prime }\beta _{0}+\alpha _{g})$ and define the $2^{T}\times 2^{T}$
matrix $\breve{P}$ with $\breve{P}_{g,h}=P_{g,T}^{y_{T}}\prod%
\limits_{t=1}^{T-1}(P_{g,t}(1+P_{g,t+1})/(1+P_{g,t+1}e^{\gamma
_{0}}))^{y_{t}}$ for $g,h=1,2,...,2^{T}$ with $%
h=1+2^{0}y_{1}+2^{1}y_{2}+...+2^{T-1}y_{T}.$ Finally, let $\overline{D}=%
\overline{D}(x,\beta _{0},\gamma _{0},\alpha )$ and $\breve{D}=\breve{D}%
(\gamma _{0})$ be nonsingular diagonal matrices with $\overline{D}%
_{g,g}=((1+P_{g,1}e^{\gamma
_{0}})/(1+P_{g,1}))^{y_{0}}\prod\limits_{t=1}^{T}(1+P_{g,t})$ and $\breve{D}%
_{h,h}=\prod\limits_{t=1}^{T}\exp (-\gamma _{0}y_{t-1}y_{t})$ for $%
g,h=1,2,...,2^{T}$ with $h=1+2^{0}y_{1}+2^{1}y_{2}+...+2^{T-1}y_{T}.$ Then
it is easily verified that $\breve{P}=\overline{D}\bar{P}\breve{D}.$ Hence $%
rk(\breve{P})=rk(\bar{P}).$ We also define $y^{S}=\tsum%
\nolimits_{t=1}^{T}y_{t}$ for later use.

We now show (i). If the model does not contain covariates, i.e., if $\beta
_{0}=0,$ then $\breve{P}$ does not depend on $x$ and there exist $2^{T}-rk(%
\breve{P})$ linearly independent moment functions, which will not depend on $%
x$. Furthermore, the number of linearly independent moment functions
available for the model without covariates is at least as large as the
number of linearly independent moment functions available for the model that
does include them, i.e., that allows $\beta _{0}\neq 0$. In the appendix we
show that $rk(\breve{P})\geq 2T$ irrespective of whether $\beta _{0}=0$ or $%
\beta _{0}\neq 0$, that is, we prove Lemma 1, which states that the $2T$
columns of $\breve{P}$ corresponding to vectors $y$ with either the first $k$
or the last $k$ elements equal to 1 and the remaining elements (if any)
equal to 0 for $k=0,1,2,...T$ are linearly independent.\thinspace \footnote{%
More generally, any $2T$ columns of $\breve{P}$ will be linearly independent
if they correspond to the following $2T$ $y$-vectors: the two $y$-vectors
that satisfy $y^{S}=0$ or $y^{S}=T$ and for each $k\in $ $\{1,2,...,T-1\}$
two $y$-vectors that satisfy $y^{S}=k$, one with $y_{T}=0$ and the other
with $y_{T}=1.$} Recall that $rk(\breve{P})=rk(\bar{P}).$ It follows that
claim (i) is correct. We now show (ii):

It is easily seen that for any $t_{1}$ and $s_{1}$ with $t_{1}<s_{1}<T$, the 
$2^{t_{1}}$ moment functions\linebreak $%
w_{y_{1},...,y_{t_{1}-1}}(y_{1},...,y_{t_{1}-1})m_{y_{0}}^{(a/b)(t_{1},s_{1},T)} 
$ are linearly independent because the $2^{t_{1}-1}$ indicator functions $%
w_{y_{1},...,y_{t_{1}-1}}(y_{1},...,y_{t_{1}-1})$ are linearly independent
and $m_{y_{0}}^{(a)(t_{1},s_{1},T)}$ and $m_{y_{0}}^{(b)(t_{1},s_{1},T)}$
are linearly independent. Furthermore, any nontrivial linear combination of
the moment functions $%
w_{y_{1},...,y_{t_{1}-1}}(y_{1},...,y_{t_{1}-1})m_{y_{0}}^{(a/b)(t_{1},s_{1},T)}(y,x,\beta ,\gamma ) 
$ with $t_{1}<s_{1}<T$ is linearly independent of $%
w_{y_{1},...,y_{t-1}}(y_{1},...,y_{t-1})m_{y_{0}}^{(a/b)(t,s,T)}(y,x,\beta
,\gamma )$ with $t<s<T$ and $(t,s)\neq (t_{1},s_{1})$\linebreak because only
the former depends on $\exp [\pm z_{t_{1},s_{1}}(y_{0},y,x,\beta ,\gamma )],$
where $z_{t_{1},s_{1}}(y_{0},y,x,\beta ,\gamma
)=(x_{t_{1}}-x_{s_{1}})^{\prime }\beta +\gamma (y_{t_{1}-1}-y_{s_{1}-1})$.
This is still true when $\beta =0.$ Hence the $2^{T}-2T$ functions $%
w_{y_{1},...,y_{t-1}}(y_{1},...,y_{t-1})m_{y_{0}}^{(a/b)(t,s,T)}(y,x,\beta
,\gamma )$ are linearly independent. They are also valid moment functions,
see Honor\'{e} and Weidner (2020). It follows that they span a $2^{T}-2T$%
\thinspace -dimensional linear subspace of the $2^{T}$-dimensional vector
space of real valued functions over the outcomes $y\in \{0,1\}^{T}$ that
contains the valid moment functions.

Remark 1: The analysis above is also valid when there are no covariates,
i.e., $\beta _{0}=0$.

Remark 2: When $\beta _{0}\neq 0,$ then $\breve{P}$ depends on $x$ and part
(i) of the proof implies that $rk(\breve{P})\geq 2T.$ However, part (ii) of
the proof shows that there exist at least $2^{T}-2T$ linearly\linebreak
independent moment functions, which in turn implies that $rk(\breve{P})\leq
2T$. We conclude that $rk(\breve{P})=2T.$ When $\beta _{0}=0,$ the proof of
the conjecture still implies that $rk(\breve{P})=2T$.

Remark 3: It follows from the result under (i) that there are no valid
moment functions when $T=2.$ In other words, GMM estimation of the panel
logit AR(1) model with fixed effects and possibly strictly exogenous
covariates is not possible for $T=2.$ Our proof is more general than that of
Honor\'{e} and Weidner (2020) for this claim because we also cover the case
where the values of $\alpha $ can only be finite. In their proof, Honor\'{e}
and Weidner (2020) chose two of the four different values of $\alpha $ equal
to $\pm \infty ,$ which leads to probabilities that are equal to 1 for the
events where all elements of $y$ are either zero or one. This unnecessarily
restricts the moment functions a priori. In contrast, we also allow all the
probabilities of observing a $y$-vector with only zeros or only ones to be
less than 1.

Remark 4: The analysis above can also be extended to panel logit AR($p$)
models with fixed effects and $p>1$.

Remark 5: The analysis above can also be used for the static panel logit
model, i.e., when $\gamma _{0}=0.$ In that case $\breve{P}%
_{g,h}=\prod\limits_{t=1}^{T}P_{g,t}^{y_{t}}.$ When also $\beta _{0}=0,$ $%
\breve{P}$ is equal to a matrix with columns from a Vandermonde matrix of
rank $T+1.$ It follows that when $\gamma _{0}=0,$ the set of valid moment
functions is a linear subspace of the $2^{T}$-dimensional vector space of
real valued functions over the outcomes $y\in \{0,1\}^{T}$ that has at most
dimension $2^{T}-(T+1)$ and in particular that when $T=2,$ there exists at
most one valid moment condition.

\section{Some results for the panel logit AR(2) model}

When $p=2,$ we have%
\begin{eqnarray*}
Pr(Y_{i} &=&y_{i}|Y_{i0}=y_{i0},Y_{i,-1}=y_{i,-1},\text{ }%
X_{i}=x_{i},A_{i}=\alpha _{i})\equiv \\
p_{y_{i}^{(0)}}(y_{i},x_{i},\beta _{0},\gamma _{0},\alpha _{i})
&=&\prod\limits_{t=1}^{T}\frac{exp(x_{it}^{\prime }\beta
_{0}+\tsum\nolimits_{l=1}^{2}y_{i,t-l}\gamma _{l,0}+\alpha _{i})}{%
1+exp(x_{it}^{\prime }\beta _{0}+\tsum\nolimits_{l=1}^{2}y_{i,t-l}\gamma
_{l,0}+\alpha _{i})},
\end{eqnarray*}%
where $y_{i}^{(0)}=(y_{i0},$ $y_{i,-1})^{\prime }$ and $\gamma _{0}=(\gamma
_{1,0},\gamma _{2,0})^{\prime }.$ We drop the index $i.$ Let us redefine $%
\bar{P}$ as a $2^{T}\times 2^{T}$ matrix with $\bar{P}%
_{g,h}=p_{y^{(0)}}(y,x,\beta _{0},\gamma _{0},\alpha _{g})$ for $%
g,h=1,2,...,2^{T}$ with $h=1+2^{0}y_{1}+2^{1}y_{2}+...+2^{T-1}y_{T},$ and
let us redefine $\breve{P}$ as a $2^{T}\times 2^{T}$ matrix with $\breve{P}%
_{g,h}=P_{g,T}^{y_{T}}\prod\limits_{t=2}^{T-1}\left( P_{g,t-1}(\frac{%
(1+P_{g,t})(1+P_{g,t+1})}{(1+P_{g,t}e^{\gamma _{1}})(1+P_{g,t+1}e^{\gamma
_{2}})})^{1-y_{t-2}}(\frac{1+P_{g,t}}{1+P_{g,t}e^{\gamma _{1}+}{}^{\gamma
_{2}}})^{y_{t-2}}\right) ^{y_{t-1}}\times \newline
\left( P_{g,T-1}(\frac{1+P_{g,T}}{1+P_{g,T}e^{\gamma _{1}}})^{1-y_{T-2}}(%
\frac{1+P_{g,T}}{1+P_{g,T}e^{\gamma _{1}+}{}^{\gamma _{2}}}%
)^{y_{T-2}}\right) ^{y_{T-1}}$ for $g,h=1,2,...,2^{T}$ with $%
h=1+2^{0}y_{1}+2^{1}y_{2}+...+2^{T-1}y_{T}.$ Note that with these new
definitions of $\bar{P}$ and $\breve{P},$ we still have $\breve{P}=\overline{%
D}\bar{P}\breve{D}$ for some nonsingular diagonal matrices $\overline{D}=%
\overline{D}(x,\beta _{0},\gamma _{0},\alpha )$ and $\breve{D}=\breve{D}%
(\gamma _{0})$.

The formula for $\breve{P}_{g,h}$ suggests that a second conjecture of Honor%
\'{e} and Weidner (2020), henceforth H\&W, namely that the number of
linearly independent moment functions for the general panel logit AR($p$)
models with covariates is given by $l=2^{T}-(T-p+1)2^{p},$ is plausible:
when $p$ increases by one, the number of possible values for a $p$-tuple $%
(y_{t-p},y_{t-p+1},\ldots ,y_{t-1}),$ namely $2^{p},$ doubles, while the
number of different sets of $p$ consecutive elements of $%
\{y_{1},y_{2},...,y_{T-1}\}$ that appear in the products of powers in $%
\breve{P}_{g,h}$ decreases by one (this number equals $T-2$ when $p=2$) and
the factors in $\breve{P}_{g,h}$ whose power depends on either $y_{T}$ or $%
y_{0}$ account for $2^{p}$ more possibilities, which explains the $(T-p+1)$
part of the formula. To prove H\&W's second conjecture for $p>1,$ one can in
principle follow a similar proof strategy as for the case where $p=1$.
However, when $p>1,$ things are a bit more complicated. As H\&W demonstrate,
when $p>1,$ the number of linearly independent moment functions for the
general panel logit AR($p$) model is smaller than the number of linearly
independent moment functions for the panel logit AR($p$) model without
covariates (i.e., with $\beta _{0}=0$). One can relatively easily establish
the latter number for different values of $T$ by using a proof strategy
similar to that for the case $p=1.$ The difference between the two numbers
of moment functions is equal to the number of linearly independent "special"
moment functions that are only valid for "special" versions of the model,
e.g. the model with $\beta _{0}=0,$ but not for the general model. Thus by
subtracting the number of these special moment functions from the total
number of linearly independent moment functions for the model with $\beta
_{0}=0,$ one obtains the number of linearly independent moment functions for
the general model.

H\&W claim that they have found all moment functions for the general model
when $T\leq 5$. However, their claim is premature as they have not shown
that there cannot be more than $l$ moment functions for the general model
when $T\leq 5$.\thinspace \footnote{%
H\&W have found one moment function for the panel logit AR(2) model with $%
\beta _{0}=0$ (given the value of $y^{(0)}$) when $T=3$, which is a
"special" moment function that is only valid when $x_{2}=x_{3}.$ However,
they have not shown that when $T=3$, there is \textit{only} one moment
function for this model.} We have shown this above for $p=1$ (and any $T$)
and we will show this in the appendix for $p=2$ and $T\leq 5.$

For the panel logit AR(2) model without covariates (i.e., with $\beta _{0}=0$%
), one can show that $rk(\breve{P})=4(T-1)-(T-2)=3T-2,$ so that there are $%
2^{T}-(3T-2)$ linearly independent moment functions available for this
model.\thinspace \footnote{%
A proof strategy for the claim that $rk(\breve{P})=3T-2$ is discussed in the
appendix.} One can easily obtain these by solving the system $\bar{P}%
_{[3T-2]}\bar{M}_{3T-2}=0,$ where $\bar{P}_{[3T-2]}=\bar{P}%
_{[3T-2]}(e^{\gamma _{1,0}},$ $e^{\gamma _{2,0}})$ is a $(3T-2)\times 2^{T}$
matrix that consists of (any) $3T-2$ rows of the matrix $\bar{P}$, each
evaluated at/corresponding to different values for the $\alpha _{g},$ and $%
\bar{M}_{3T-2}$ is a $2^{T}\times (2^{T}-(3T-2))$ matrix with $rk(\bar{M}%
_{3T-2})=2^{T}-(3T-2)$. The $2^{T}-(3T-2)$ columns of $\bar{M}_{3T-2}$ span
the nullspace of $\bar{P},$ which is the space of valid moment functions for
the panel logit AR(2) model without covariates.

\appendix

\section{Appendix}

\begin{lemma}
The $2T$ columns of $\breve{P}$ corresponding to vectors $y$ with either the
first $k$ or the last $k$ elements equal to 1 and the remaining elements (if
any) equal to 0 for $k=0,1,2,...T$ are linearly independent \textit{a.s.}
(almost surely) for any $T\geq 2$:
\end{lemma}

\textbf{Proof: }We will prove this Lemma by showing that the square matrix $%
\widetilde{P}_{2T}$ (sometimes denoted by $\widetilde{P}$ for short for some
value of $T$) that contains the first $2T$ rows of these $2T$ columns of $%
\breve{P}$ has full rank for any $T\geq 2$. We will omit the subscript $0$
from $\beta _{0}$ and $\gamma _{0}$.

We will first consider the special (and most challenging) case where $\beta
=0.$

We define the elements of the matrix $\widetilde{P}_{2T}$ as follows:

If $y=(1,\ldots ,1,0,\ldots ,0)^{\prime }$ with the first $k$ entries equal
to 1 and $0\leq k\leq T-1:$

$\widetilde{P}_{2T,g,h}$ (or simply $\widetilde{P}_{g,h}$ for some value of $%
T$) $=(e^{\alpha _{g}}\frac{1+e^{\alpha _{g}}}{1+e^{\alpha _{g}+\gamma }}%
)^{k}$ for any $g\in \{1,2,\ldots ,2T\}$ and for $h=2k+1$;

if $y=(0,\ldots ,0,1,\ldots ,1)^{\prime }$ with the last $k+1$ entries equal
to 1 and $0\leq k\leq T-1:$

$\widetilde{P}_{2T,g,h}=(\widetilde{P}_{g,h}=)$ $e^{\alpha _{g}}(e^{\alpha
_{g}}\frac{1+e^{\alpha _{g}}}{1+e^{\alpha _{g}+\gamma }})^{k}$ for any $g\in
\{1,2,\ldots ,2T\}$ and for $h=2(k+1).$

Let $D_{2T}=diag(1+e^{\alpha _{1}+\gamma },1+e^{\alpha _{2}+\gamma },\ldots
,1+e^{\alpha _{2T}+\gamma }).$ Note that $\det (D_{2T})\neq 0.$

We will prove the Lemma by induction.\smallskip\ When $T=2,$ we consider the 
$4\times 4$\ matrix\linebreak $D_{4}\widetilde{P}_{4}=%
\begin{bmatrix}
1+e^{\alpha _{1}+\gamma } & e^{\alpha _{1}}(1+e^{\alpha _{1}+\gamma }) & 
e^{\alpha _{1}}(1+e^{\alpha _{1}}) & e^{2\alpha _{1}}(1+e^{\alpha _{1}}) \\ 
1+e^{\alpha _{2}+\gamma } & e^{\alpha _{2}}(1+e^{\alpha _{2}+\gamma }) & 
e^{\alpha _{2}}(1+e^{\alpha _{2}}) & e^{2\alpha _{2}}(1+e^{\alpha _{2}}) \\ 
1+e^{\alpha _{3}+\gamma } & e^{\alpha _{3}}(1+e^{\alpha _{3}+\gamma }) & 
e^{\alpha _{3}}(1+e^{\alpha _{3}}) & e^{2\alpha _{3}}(1+e^{\alpha _{3}}) \\ 
1+e^{\alpha _{4}+\gamma } & e^{\alpha _{4}}(1+e^{\alpha _{4}+\gamma }) & 
e^{\alpha _{4}}(1+e^{\alpha _{4}}) & e^{2\alpha _{4}}(1+e^{\alpha _{4}})%
\end{bmatrix}%
,\smallskip $ and it is easily verified that $rank(D_{4}\widetilde{P}_{4})=4$
a.s. (Recall that $\gamma \neq 0$, note that any linear combination of the
first two columns of $D_{4}\widetilde{P}_{4}$ depends on $\gamma $ and
conclude that the k-th column of $D_{4}\widetilde{P}_{4}$ cannot be written
as a linear combination of the k-1 columns on its LHS for $k=2,\ldots ,4$).
As $\det (D_{4})\neq 0,$ it follows that $rank(\widetilde{P}_{4})=4$ a.s.

Assuming that the Lemma is correct for $T=S+2$ for some $S\in 
\mathbb{N}
,$ we will now prove that it is also correct for $T=S+3:$

The $2(S+3)\times 2(S+3)$ matrix $\widetilde{P}=\widetilde{P}_{2(S+3)}$
contains the $2(S+2)\times 2(S+2)$ matrix $\widetilde{P}_{2(S+2)}$ (in the
north-west corner) and two more rows and columns:

\noindent $\widetilde{P}_{2(S+3)}=\left[ 
\begin{tabular}{ccc}
$\widetilde{P}_{2(S+2)}$ & $[\breve{P}_{g,2^{S+2}}]_{1\leq g\leq 2(S+2)}$ & $%
[\breve{P}_{g,2^{S+3}}]_{1\leq g\leq 2(S+2)}$ \\ 
$\lbrack \widetilde{P}_{2S+5,h}]_{1\leq h\leq 2(S+2)}$ & $\breve{P}%
_{2S+5,2^{S+2}}$ & $\breve{P}_{2S+5,2^{S+3}}$ \\ 
$\lbrack \widetilde{P}_{2S+6,h}]_{1\leq h\leq 2(S+2)}$ & $\breve{P}%
_{2S+6,2^{S+2}}$ & $\breve{P}_{2S+6,2^{S+3}}$%
\end{tabular}%
\right] ,$ \smallskip where $\breve{P}$ is a\linebreak $2(S+3)\times 2(S+3)$
matrix.

We can partition $D_{2(S+3)}^{S+2}\widetilde{P}=D_{2(S+3)}^{S+2}\widetilde{P}%
_{2(S+3)}$ as $\left[ 
\begin{tabular}{cc}
$D_{2(S+2)}^{S+2}\widetilde{P}_{2(S+2)}$ & $B$ \\ 
$C$ & $F$%
\end{tabular}%
\right] .$\smallskip

Let $M\equiv D_{2(S+2)}^{S+2}\widetilde{P}_{2(S+2)}-BF^{-1}C.$ Then it
follows from a standard result regarding the determinants of partitioned
matrices that $\det (D_{2(S+3)}^{S+2}\widetilde{P})=\det (F)\det (M).$

It is easily checked that $F$ has full rank, i.e., $rank(F)=2$:

$F=\left[ 
\begin{tabular}{cc}
$(e^{\alpha _{2S+5}}(1+e^{\alpha _{2S+5}}))^{S+2}$ & $e^{\alpha
_{2S+5}}(e^{\alpha _{2S+5}}(1+e^{\alpha _{2S+5}}))^{S+2}$ \\ 
$(e^{\alpha _{2S+6}}(1+e^{\alpha _{2S+6}}))^{S+2}$ & $e^{\alpha
_{2S+6}}(e^{\alpha _{2S+6}}(1+e^{\alpha _{2S+6}}))^{S+2}$%
\end{tabular}%
\right] $\smallskip\ so $\det (F)\neq 0$ because $e^{\alpha
_{2S+6}}-e^{\alpha _{2S+5}}\neq 0.$\smallskip

It is also easily shown that $M=D_{2(S+2)}^{S+2}\widetilde{P}%
_{2(S+2)}-BF^{-1}C$ is invertible because it follows from Leibniz's formula
for determinants (or from Laplace's expansion of the determinant, which uses
cofactors and minors) that $\det (M)$ is equal to a polynomial in the
elements of $M,$ because this polynomial can be rewritten as a sum of terms
that includes the term $\det (D_{2(S+2)}^{S+2}\widetilde{P}_{2(S+2)}),$
because $\det (D_{2(S+2)}^{S+2}\widetilde{P}_{2(S+2)})\neq 0$ a.s., and
because (the sum of) all the other terms in this sum is/are a.s. incapable
of canceling out $\det (D_{2(S+2)}^{S+2}\widetilde{P}_{2(S+2)})$:

Let $Q=BF^{-1}C\equiv \widetilde{Q}/\det (F).$ Then $%
Q_{g,h}=B_{g,.}F^{-1}C_{.,h}=\widetilde{Q}_{g,h}/\det (F)$ with $\widetilde{Q%
}_{g,h}=$\smallskip \linebreak $(1+e^{\alpha _{g}+\gamma })^{S+2}%
\begin{bmatrix}
\breve{P}_{g,2^{S+2}} & \breve{P}_{g,2^{S+3}}%
\end{bmatrix}%
\left[ 
\begin{tabular}{cc}
$e^{\alpha _{2S+6}}(e^{\alpha _{2S+6}}(1+e^{\alpha _{2S+6}}))^{S+2}$ & $%
-e^{\alpha _{2S+5}}(e^{\alpha _{2S+5}}(1+e^{\alpha _{2S+5}}))^{S+2}$ \\ 
$-(e^{\alpha _{2S+6}}(1+e^{\alpha _{2S+6}}))^{S+2}$ & $(e^{\alpha
_{2S+5}}(1+e^{\alpha _{2S+5}}))^{S+2}$%
\end{tabular}%
\right] \times $\smallskip \linebreak $\left[ 
\begin{array}{c}
\widetilde{P}_{2S+5,h}(1+e^{\alpha _{2S+5}})^{S+2} \\ 
\widetilde{P}_{2S+6,h}(1+e^{\alpha _{2S+6}})^{S+2}%
\end{array}%
\right] =%
\begin{bmatrix}
(e^{\alpha _{g}}(1+e^{\alpha _{g}}))^{S+2} & e^{\alpha _{g}}(e^{\alpha
_{g}}(1+e^{\alpha _{g}}))^{S+2}%
\end{bmatrix}%
\times $\smallskip \newline
$\left[ 
\begin{tabular}{cc}
$e^{\alpha _{2S+6}}(e^{\alpha _{2S+6}}(1+e^{\alpha _{2S+6}}))^{S+2}$ & $%
-e^{\alpha _{2S+5}}(e^{\alpha _{2S+5}}(1+e^{\alpha _{2S+5}}))^{S+2}$ \\ 
$-(e^{\alpha _{2S+6}}(1+e^{\alpha _{2S+6}}))^{S+2}$ & $(e^{\alpha
_{2S+5}}(1+e^{\alpha _{2S+5}}))^{S+2}$%
\end{tabular}%
\right] \times $\smallskip \linebreak $\left[ 
\begin{array}{c}
(e^{\alpha _{2S+5}})^{\delta }(e^{\alpha _{2S+5}}\frac{1+e^{\alpha _{2S+5}}}{%
1+e^{\alpha _{2S+5}+\gamma }})^{k}(1+e^{\alpha _{2S+5}})^{S+2} \\ 
(e^{\alpha _{2S+6}})^{\delta }(e^{\alpha _{2S+6}}\frac{1+e^{\alpha _{2S+6}}}{%
1+e^{\alpha _{2S+6}+\gamma }})^{k}(1+e^{\alpha _{2S+6}})^{S+2}%
\end{array}%
\right] $ for some $k\in \{0,1,\ldots ,S+1\}$ and\smallskip\ some $\delta
\in \{0,1\}.$ Omitting the factor $(e^{\alpha _{g}}(1+e^{\alpha
_{g}})(1+e^{\alpha _{2S+5}})(1+e^{\alpha _{2S+6}}))^{S+2},$\smallskip 
\begin{gather*}
\widetilde{Q}_{g,h}\propto 
\begin{bmatrix}
1 & e^{\alpha _{g}}%
\end{bmatrix}%
\begin{bmatrix}
e^{\alpha _{2S+6}}(e^{\alpha _{2S+6}})^{S+2} & -e^{\alpha _{2S+5}}(e^{\alpha
_{2S+5}})^{S+2} \\ 
-(e^{\alpha _{2S+6}})^{S+2} & (e^{\alpha _{2S+5}})^{S+2}%
\end{bmatrix}%
\begin{bmatrix}
(e^{\alpha _{2S+5}})^{\delta }(e^{\alpha _{2S+5}}\frac{1+e^{\alpha _{2S+5}}}{%
1+e^{\alpha _{2S+5}+\gamma }})^{k} \\ 
(e^{\alpha _{2S+6}})^{\delta }(e^{\alpha _{2S+6}}\frac{1+e^{\alpha _{2S+6}}}{%
1+e^{\alpha _{2S+6}+\gamma }})^{k}%
\end{bmatrix}%
= \\
e^{\delta \alpha _{2S+6}}\left( e^{\alpha _{g}}e^{(S+2)\alpha
_{2S+5}}-e^{\alpha _{2S+5}}e^{(S+2)\alpha _{2S+5}}\right) \left( \frac{%
e^{\alpha _{2S+6}}}{e^{\gamma +\alpha _{2S+6}}+1}\left( e^{\alpha
_{2S+6}}+1\right) \right) ^{k}- \\
e^{\delta \alpha _{2S+5}}\left( e^{\alpha _{g}}e^{(S+2)\alpha
_{2S+6}}-e^{\alpha _{2S+6}}e^{(S+2)\alpha _{2S+6}}\right) \left( \frac{%
e^{\alpha _{2S+5}}}{e^{\gamma +\alpha _{2S+5}}+1}\left( e^{\alpha
_{2S+5}}+1\right) \right) ^{k}= \\
e^{\delta \alpha _{2S+6}}e^{(S+2)\alpha _{2S+5}}\left( e^{\alpha
_{g}}-e^{\alpha _{2S+5}}\right) \left( \frac{e^{\alpha _{2S+6}}}{e^{\gamma
+\alpha _{2S+6}}+1}\left( e^{\alpha _{2S+6}}+1\right) \right) ^{k}- \\
e^{\delta \alpha _{2S+5}}e^{(S+2)\alpha _{2S+6}}\left( e^{\alpha
_{g}}-e^{\alpha _{2S+6}}\right) \left( \frac{e^{\alpha _{2S+5}}}{e^{\gamma
+\alpha _{2S+5}}+1}\left( e^{\alpha _{2S+5}}+1\right) \right) ^{k}.
\end{gather*}%
Note that the expression for $\widetilde{Q}_{g,h}$ cannot be rewritten as an
expression that is divisible by the expression $e^{\alpha _{2S+6}}-e^{\alpha
_{2S+5}}$ and hence that the expressions for all elements of $Q$ are ratios
with the factor $e^{\alpha _{2S+6}}-e^{\alpha _{2S+5}}$ in the denominator.
We conclude that $\det (M)$ can be written as the sum of $\det
(D_{2(S+2)}^{S+2}\widetilde{P}_{2(S+2)})$ and one other term, (which itself
is the result of summing almost all terms that appear in the aforementioned
expansion of $\det (M)$ except for $\det (D_{2(S+2)}^{S+2}\widetilde{P}%
_{2(S+2)}),$ and) which is an expression that is given by a ratio with the
factor $e^{\alpha _{2S+6}}-e^{\alpha _{2S+5}}$ raised to some positive power
appearing in the denominator (as a common factor) and with the same factor
also appearing in the numerator but raised to lower positive powers than its
power in the denominator so that its presence in the numerator does not
completely cancel out this factor in the denominator.\thinspace \footnote{%
We have not investigated whether this second term (expression) in the sum is
zero. If the latter were the case, we would have $\det (M)=\det
(D_{2(S+2)}^{S+2}\widetilde{P}_{2(S+2)})\neq 0$ a.s., i.e., $\det (M)\neq 0$
a.s., which is what we want to show.} However, none of the elements of $%
D_{2(S+2)}^{S+2}\widetilde{P}_{2(S+2)}$ depend on $e^{\alpha _{2S+5}}$ or $%
e^{\alpha _{2S+6}}.$ It follows that $\det (M)\neq 0$ a.s. and that $%
\widetilde{P}=\widetilde{P}_{2(S+3)}$ is invertible a.s. (as we have already
seen that $\det (F)\neq 0$), i.e., $rank(\widetilde{P}_{2(S+3)})=2(S+3)$
a.s. Another way of seeing this is that $\det (M)$ can be expressed as a
ratio with a numerator that is a polynomial in $e^{\alpha _{g}}$ for $%
g=1,2,\ldots ,2(S+2),$\ in $e^{\gamma }$ and, unless the second term ("the
other term") in the aforementioned sum of two terms is zero (in which case $%
\det (M)=\det (D_{2(S+2)}^{S+2}\widetilde{P}_{2(S+2)})\neq 0$ a.s.), also in 
$e^{\alpha _{2S+5}}$ and $e^{\alpha _{2S+6}}.$ Hence $\det (M)=0$ if and
only if this numerator equals zero. Given values of $e^{\alpha _{g}}$ for $%
g=1,2,\ldots ,2(S+2)$ and $e^{\gamma },$ the numerator is a polynomial in $%
e^{\alpha _{2S+5}}$ and $e^{\alpha _{2S+6}}$ with a finite number of roots.
As the values of $\alpha _{g},$ $g=1,2,\ldots ,2(S+3)$, and $\gamma \neq 0$
can be assumed to be randomly drawn from some continuous distribution(s),
the probability that the values of $e^{\alpha _{2S+5}}$ and $e^{\alpha
_{2S+6}}$ coincide with these roots is negligible. It follows that $\Pr
(\det (M)\neq 0)=1$ and hence that $\Pr (\det (\widetilde{P}_{2(S+3)})\neq
0)=1.$

The arguments generalize to the case where $\beta \neq 0.$

Q.E.D.\bigskip

\emph{An alternative proof of the claim that }$rk(\bar{P})=2T$\emph{\ for
the panel logit AR(1) model without covariates (i.e., with }$\beta _{0}=0$%
\emph{):\medskip }

Consider the $2^{T}\times 2^{T}$ matrix $\ddot{P}$ with typical element $%
\ddot{P}_{g,h}=(1+P_{g,1}e^{\gamma })^{T-1}P_{g,1}^{y^{S}}\times $ $%
\prod\limits_{t=1}^{T-1}((1+P_{g,1})/(1+P_{g,1}e^{\gamma }))^{y_{t}}$ for $%
g,h=1,2,...,2^{T}$ with $h=1+2^{0}y_{1}+2^{1}y_{2}+...+2^{T-1}y_{T}.$ Note
that $\ddot{P}=\ddot{D}\breve{P}$ for some nonsingular diagonal matrix $%
\ddot{D}=\ddot{D}(\gamma _{0},\alpha )$ and that the columns of $\ddot{P}$
correspond to different polynomials in $P_{g,1}$ up to order $2T-1$ with all
intermediate powers occuring somewhere inside $\ddot{P}$. It follows that $%
rk(\bar{P})=rk(\ddot{P})$ is equal to the rank of a matrix that consists of
linear combinations of the columns of a Vandermonde matrix that is based on
powers of $P_{g,1}$ and has rank $2T.$ Hence $rk(\bar{P})=rk(\ddot{P})\leq
2T.$ To prove that $rk(\bar{P})=rk(\ddot{P})=2T,$ it suffices to show that $%
rk(\ddot{P})\geq 2T.$ This can be done by selecting the same $2T$ columns of 
$\ddot{P}$\ as those of $\breve{P}$ that underlie the definition of the
matrix $\widetilde{P}$ that is used in the proof of Lemma 1. Of course, it
follows from Lemma 1, $rk(\ddot{D})=2^{T}$ and $\ddot{P}=\ddot{D}\breve{P}$
that $rk(\ddot{P})\geq 2T.$\bigskip

\textbf{Analysis for the panel logit AR(2) model:}\medskip

\emph{Proof strategy for the claim that }$rk(\bar{P})=3T-2$\emph{\ for the
panel logit AR(2) model without covariates (i.e., with }$\beta _{0}=0$\emph{%
):\medskip }

When $y_{0}=1,$ we consider $\ddot{P}$ with typical element $\ddot{P}%
_{g,h}=(1+P_{g,1}e^{\gamma _{1}})^{\left\lfloor 0.5(T-1)\right\rfloor
}(1+P_{g,1}e^{\gamma _{2}})^{\left\lfloor 0.5(T-2)\right\rfloor
}(1+P_{g,1}e^{\gamma _{1}+}{}^{\gamma _{2}})^{T-1}P_{g,1}^{y^{S}}\left( (%
\frac{1+P_{g,1}}{1+P_{g,1}e^{\gamma _{1}}})^{1-y_{T-2}}(\frac{1+P_{g,1}}{%
1+P_{g,1}e^{\gamma _{1}+}{}^{\gamma _{2}}})^{y_{T-2}}\right)
^{y_{T-1}}\times \newline
\prod\limits_{t=2}^{T-1}\left( (\frac{(1+P_{g,1})^{2}}{(1+P_{g,1}e^{\gamma
_{1}})(1+P_{g,1}e^{\gamma _{2}})})^{1-y_{t-2}}(\frac{1+P_{g,1}}{%
1+P_{g,1}e^{\gamma _{1}+}{}^{\gamma _{2}}})^{y_{t-2}}\right) ^{y_{t-1}}.$
Note that $\ddot{P}=\ddot{D}\breve{P}$ for some nonsingular diagonal matrix $%
\ddot{D}=\ddot{D}(\gamma _{0},\alpha )$ and that the columns of $\ddot{P}$
correspond to different polynomials in $P_{g,1}$ up to order $3(T-1)$ with
all intermediate powers occuring somewhere inside $\ddot{P}$. It follows
that $rk(\bar{P})=rk(\ddot{P})$ is equal to the rank of a matrix that
consists of linear combinations of the columns of a Vandermonde matrix that
is based on powers of $P_{g,1}$ and has rank $3T-2.$ Hence $rk(\bar{P})=rk(%
\ddot{P})\leq 3T-2.$ To prove that $rk(\bar{P})=rk(\ddot{P})=3T-2,$ it
suffices to show that $rk(\ddot{P})\geq 3T-2.$ This can be done by selecting 
$3T-2$ suitable columns of $\ddot{P}$\ and showing that they are linearly
independent similarly to the proof of Lemma 1.

When $y_{0}=0,$ we consider $\ddot{P}$ with typical element $\ddot{P}%
_{g,h}=(1+P_{g,1}e^{\gamma _{1}})^{\left\lfloor 0.5T\right\rfloor
}(1+P_{g,1}e^{\gamma _{2}})^{\left\lfloor 0.5(T-1)\right\rfloor
}(1+P_{g,1}e^{\gamma _{1}+}{}^{\gamma _{2}})^{T-2}P_{g,1}^{y^{S}}\left( (%
\frac{1+P_{g,1}}{1+P_{g,1}e^{\gamma _{1}}})^{1-y_{T-2}}(\frac{1+P_{g,1}}{%
1+P_{g,1}e^{\gamma _{1}+}{}^{\gamma _{2}}})^{y_{T-2}}\right)
^{y_{T-1}}\times \newline
\prod\limits_{t=2}^{T-1}\left( (\frac{(1+P_{g,1})^{2}}{(1+P_{g,1}e^{\gamma
_{1}})(1+P_{g,1}e^{\gamma _{2}})})^{1-y_{t-2}}(\frac{1+P_{g,1}}{%
1+P_{g,1}e^{\gamma _{1}+}{}^{\gamma _{2}}})^{y_{t-2}}\right) ^{y_{t-1}}.$
Note that $\ddot{P}=\ddot{D}\breve{P}$ for some nonsingular diagonal matrix $%
\ddot{D}=\ddot{D}(\gamma _{0},\alpha )$ and that the columns of $\ddot{P}$
correspond to different polynomials in $P_{g,1}$ up to order $3(T-1)$ with
all intermediate powers occuring somewhere inside $\ddot{P}$. It follows
that $rk(\bar{P})=rk(\ddot{P})$ is equal to the rank of a matrix that
consists of linear combinations of the columns of a Vandermonde matrix that
is based on powers of $P_{g,1}$ and has rank $3T-2.$ Hence $rk(\bar{P})=rk(%
\ddot{P})\leq 3T-2.$ To prove that $rk(\bar{P})=rk(\ddot{P})=3T-2,$ it
suffices to show that $rk(\ddot{P})\geq 3T-2.$ This can be done by selecting 
$3T-2$ suitable columns of $\ddot{P}$\ and showing that they are linearly
independent similarly to the proof of Lemma 1.\bigskip

\emph{Proof of the second conjecture of H\&W (2020) for }$p=2$\emph{\ and }$%
T\in \{3,4,5\}$:\emph{\medskip }

We have followed the proof strategy discussed above to show that when $p=2$
and $\beta _{0}=0,$ then $rk(\bar{P})=3T-2$ for all $T\in \{3,4,5\}$ and any 
$y_{0}\in \{0,1\}.$ In particular, we have used \textit{Mathematica} to
verify that when $p=2$ and $\beta _{0}=0,$ then $rk(\ddot{P})=3T-2$ for all $%
T\in \{3,4,5\}$ and any $y_{0}\in \{0,1\}.$ We note that when $p=2,$ $T=3$
and $x_{2}=x_{3},$ there is (at least) one extra moment function relative to
the number of linearly independent moment functions for the general model
(given the value of $y^{(0)}$), cf. H\&W (2020) who found one extra moment
function for this case; by analogy, when $p=2,$ $T=4$ and $%
x_{2}=x_{3}=x_{4}, $ there will be (at least) two extra moment functions
relative to the number of linearly independent moment functions for the
general model (given the value of $y^{(0)}$); and when $p=2,$ $T=5$ and $%
x_{2}=x_{3}=x_{4}=x_{5},$ there will be (at least) three extra moment
functions relative to the number of linearly independent moment functions
for the general model (given the value of $y^{(0)}$). H\&W (2020) also found 
$l=2^{T}-4(T-1)$ linearly independent moment functions for the general model
(given the value of $y^{(0)}$) when $p=2$ and $T\in \{3,4,5\},$ so there are
at least $l$ of them in these cases. Hence the number of linearly
independent "general" and "special" moment functions is at least $%
2^{T}-4(T-1)+(T-2)=2^{T}-(3T-2).$ However, this number cannot be larger than
the number of linearly independent moment functions for the model without
covariates (i.e., with $\beta _{0}=0$), which is $2^{T}-(3T-2).$ We conclude
that when $p=2$ and $T\in \{3,4,5\},$ there are $%
2^{T}-(3T-2)-(T-2)=2^{T}-4(T-1)=l$ linearly independent moment functions for
the general model (given the value of $y^{(0)}$).

Q.E.D.

\end{document}